\documentclass[prl,reprint,superscriptaddress,aps,floatfix]{revtex4-2}
\usepackage{amsmath}
\usepackage{amsfonts}
\usepackage{graphicx}
\usepackage{physics}
\usepackage[usenames]{color}
\usepackage{braket}
\usepackage{hyperref}
\hypersetup{colorlinks=true, linkcolor=blue, citecolor=blue, urlcolor=blue}

\begin{document}


\title{Coexistence of inequivalent time-crystalline orders in a Floquet collective spin system}

\author{Shashank Mishra}
\email{shashankmishra@hri.res.in}

\author{Sayan Choudhury}
\email{sayanchoudhury@hri.res.in}

\affiliation{Harish-Chandra Research Institute, Chhatnag Road, Jhunsi, Prayagraj (Allahabad) 211019, India}
\affiliation{Homi Bhabha National Institute, Training School Complex, Anushakti Nagar, Mumbai 400094, India}

\begin{abstract}
We investigate the dynamical phases that emerge in collective spin models subjected to a spatially non-uniform periodic drive. Taking the paradigmatic Lipkin-Meshkov-Glick (LMG) model as a concrete platform, we establish that a rich landscape of dynamical phases emerges when two regions of the system are driven with different field strengths, $h_1$ and $h_2$. Remarkably, despite the `all-to-all' nature of the interactions, the system can be driven into dynamical phases characterized by distinct kinds of discrete time crystal (DTC) orders in different parts of the system. Apart from these coexisting DTCs, tuning the driving field leads to the emergence of phases where DTCs coexist with Floquet-synchronized or oscillatory phases; the former has been dubbed a chimera DTC. Finally, we demonstrate that a tunable set of global DTC phases emerges when $h_1$ and $h_2$ are proximate. Crucially, these dynamical regimes can be observed both for experimentally relevant finite-size systems and in the thermodynamic limit. Our results establish spatially structured driving as a powerful route to realize non-equilibrium phase coexistence in collective spin systems.
\end{abstract}

\maketitle


{\emph{Introduction:}} In recent years, the classification of equilibrium phases and phase transitions~\cite{yeomans1992statistical,Sachdev_2011,beekman2019introduction,kivelson2024statistical} has been extended to periodically driven (Floquet) systems~\cite{bukov2015universal,oka2019floquet,weitenberg2021tailoring,harper2020topology}.
A striking example of such a Floquet phase is the
\emph{discrete time crystal} (DTC), which exhibits discrete time-translation symmetry-breaking (TTSB)~\cite{sacha2015modeling,Khemani2016,Else2016,yao2017discrete} and consequent subharmonic response of physical observables~\cite{sacha2017time,sacha2020time,else2020discrete,zaletel2023colloquium}.
Crucially, this temporal order is stable against generic perturbations,
establishing DTCs as genuine non-equilibrium phases of matter.
This stability can originate from different mechanisms including many-body localization~\cite{Khemani2016,Else2016,yao2017discrete,von2016absolute}, many-body scars~\cite{huang2018clean,huang2022discrete,maskara2021discrete,huang2023analytical,deng2023using}, collective-spin interactions~\cite{russomanno2017floquet,lyu2020eternal,biswas2025floquet,biswas2025discrete}, Hilbert-space fragmentation~\cite{kumar2025hilbert,tang2026discrete} and
long-range interactions~\cite{machado2020long,pizzi2021higher,giachetti2023fractal}. DTCs have been experimentally realized in various platforms, including trapped
ions~\cite{Zhang2017}, nitrogen-vacancy centers~\cite{Choi2017,randall2021many},
Rydberg atom arrays~\cite{bluvstein2021controlling}, and superconducting qubit
processors~\cite{xiang2024long,frey2022realization,mi2022time}. However, most studies on DTCs so far have focused on the emergence of a global DTC order under spatially \emph{uniform} driving. This raises an important question: can spatially non-uniform driving be harnessed to realize qualitatively new forms of DTC order?

\begin{figure}[t]
\centering
\includegraphics[width=0.94\linewidth]{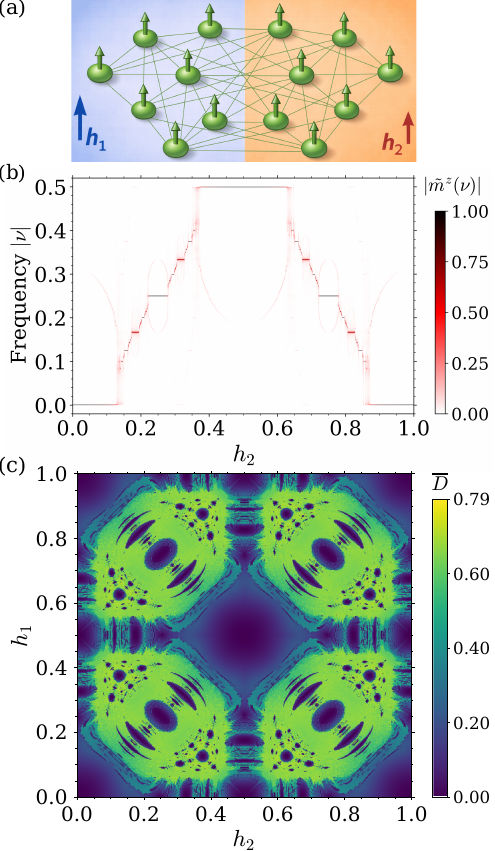}
\caption{
{\bf Model and phase diagram:} (a) Schematic of the non-uniformly driven Lipkin-Meshkov-Glick model. A globally coupled spin system is partitioned into two sub-regions of $N/2$ spins, driven by transverse fields $h_1$ and $h_2$ during the second half of the drive cycle. (b) Density plot of the Fourier transform of the stroboscopic total magnetization for the uniformly driven case ($h_1=h_2$), computed over $1000$ Floquet cycles. The system hosts a family of DTC phases including period-doubling and higher-order DTCs. (c) Density plot of $\overline{D}$, obtained by averaging $D(nT)$ (Eq.~\ref{eq:decorrelator}) from $n=1000$ to $2000$ for 1600 realizations of $\phi_1, \phi_2$. The interaction strength $J$ and the parameter $\theta$ are set to $0.5$ and $0.002$ respectively. The regions where $\overline{D}$ is small ($\sim 0$) correspond to non-ergodic dynamics, whereas large values indicate chaotic dynamics. This phase diagram identifies parameter regimes that host both global and coexisting DTC phases.}
\label{fig:1}
\end{figure}

In this Letter, we answer this question affirmatively by demonstrating that a minimal spatial inhomogeneity in the driving field induces the coexistence of distinct dynamical phases in collective spin systems. For concreteness, we investigate the dynamics of the infinite-range interacting Floquet Lipkin–Meshkov–Glick (LMG) model, in which half of the spins are driven by a transverse field $h_1$ and the remaining half by a distinct field $h_2$ [Fig.~\ref{fig:1}(a)]. This model hosts both period-doubling and higher-order global DTC phases for spatially uniform driving ($h_1 = h_2$). Interestingly, tuning $h_1$ and $h_2$ independently enables the creation of two regions that exhibit inequivalent temporal orders in a \emph{single globally coupled system}. A striking manifestation of this phenomenon is the emergence of the robust coexistence of distinct DTCs. Furthermore, this protocol enables the coexistence of a DTC phase with Floquet-synchronized or oscillatory phases; the former has been dubbed a chimera DTC~\cite{sakurai2021chimera,rahaman2024time}. Finally, we demonstrate that a tunable family of global DTCs can be realized in this system when $h_1$ and $h_2$ are comparable. Our conclusions are valid both for experimentally relevant finite system sizes and in the thermodynamic limit. Our results establish spatially structured driving as a powerful and versatile route to engineer non-equilibrium phase coexistence in collective spin systems, thereby generalizing previous studies on chimera DTCs in finite spin 
networks~\cite{sakurai2021chimera,rahaman2024time}.

{\emph{Model and stability phase diagram:}} We examine a periodically driven infinite-range LMG model, where two equal sub-regions are driven by two different transverse fields, $h_1$ and $h_2$:
\begin{align}
H(t) &=
\begin{cases}
\displaystyle
\frac{2J}{NT}
\sum_{i,j=1}^{N}
\sigma_i^z \sigma_j^z,
& 0 < t \leq \dfrac{T}{2},
\\[10pt]
\displaystyle
-\frac{2\pi}{T}
\Big(
h_1\sum_{j=1}^{N/2}\sigma_j^x
+
h_2\sum_{j=\frac{N}{2}+1}^{N}\sigma_j^x
\Big),
&
\dfrac{T}{2} < t \leq T .
\end{cases}
\label{eq:drive}
\end{align}
where $\sigma_j^{\alpha}$ ($\alpha=x,y,z$) are the standard Pauli matrices, and $J$ is the
interaction strength. A schematic illustration of this model is provided in Fig.~\ref{fig:1}(a). In the thermodynamic limit ($N\to\infty$), the dynamics of the collective-spin operators
$\hat{\mathbf S}_1=\frac{1}{2}\sum_{j=1}^{N/2}\boldsymbol{\sigma}_j$
and
$\hat{\mathbf S}_2=\frac{1}{2}\sum_{j=N/2+1}^{N}\boldsymbol{\sigma}_j$ is exactly described by semiclassical equations of motion, presented in the End Matter. When $h_1=h_2$, the system hosts both period-doubling and higher-order DTCs that exhibit robust periodic magnetization oscillations with a period, $T_{\rm DTC} > 2T$~\cite{pizzi2021higher} (see also Fig.~\ref{fig:1}(b)). We now introduce the diagnostics used to characterize the coexisting dynamical phases that emerge in this system when $h_1\ne h_2$.

\begin{figure*}[t]
\centering
\includegraphics[width=\textwidth]{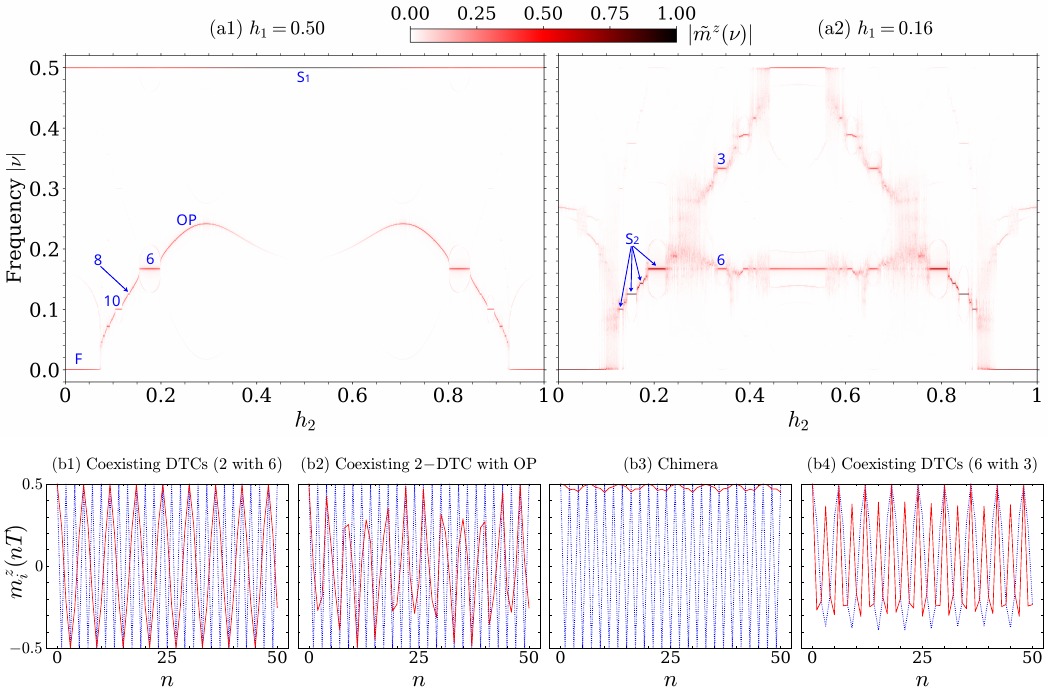}
\caption{{\bf Coexisting dynamical phases under non-uniform driving:}
Top panel: Density plots of the Fourier transform of the total magnetization, $|\tilde{m}^z(\nu)|$, along representative
cuts of the $(h_1,h_2)$ plane corresponding to $h_1=0.5$ (a1) and $h_1= 0.16$ (a2). Two simultaneous extended subharmonic plateaus at distinct frequencies signify coexisting DTCs, while a single extended subharmonic plateau corresponds to a global DTC. Panel (a1) also reveals a phase where a DTC coexists with an oscillatory phase characterized by a continuously changing Fourier spectrum. Additionally, a chimera DTC, where a DTC coexists with a Floquet-synchronized phase, is observed.
Bottom panel: The corresponding stroboscopic dynamics of the regional
magnetizations, $m_1^z$ (blue) and $m_2^z$ (red), for representative
parameter points from (a1) and (a2).
For $h_1=0.5$, coexistence between a $2T$-DTC and a
$6T$-DTC at $h_2=0.17$ (b1), coexistence between a $2T$-DTC and an
oscillatory phase at $h_2=0.25$ (b2), and a chimera DTC at
$h_2=0.03$ (b3) is observed. For $h_1=0.16$, the coexistence between a $6T$-DTC and a $3T$-DTC is shown at $h_2=0.34$ (b4). Together, these results illustrate the diverse array of coexisting dynamical phases that emerge due to spatially structured driving, despite the infinite-range nature of the interactions.
}
\label{fig:2}
\end{figure*}

The primary observable that we analyze is the subregion magnetization, ${\bf m}_i = \frac{2}{N} \langle\hat{\mathbf S}_i \rangle$, and we take a two-step approach to establish and characterize dynamical phase coexistence. Firstly, we determine the parameter regimes where the system exhibits non-ergodic behavior by computing the decorrelator, which captures the sensitivity to initial conditions:
\begin{equation}
D(nT)= \sqrt{\frac{1}{2}\sum_{i=1}^{2} \left|\mathbf m_i(nT)-\mathbf m^{\prime}_i(nT)\right|^2},
\label{eq:decorrelator}
\end{equation}
where 
$\mathbf m_i(nT)$ and $\mathbf m^{\prime}_i(nT)$ denote the magnetization after $n$ drive periods for two initially close copies of the initial state. The non-ergodic and chaotic regimes are characterized by small ($\sim 0$) and large ($\sim 1/\sqrt{2}$) values of $D(nT)$ respectively~\cite{PhysRevLett.127.140602}. The decorrelator thus serves as a period-agnostic tool to detect non-ergodic dynamical phases such as DTCs~\cite{giachetti2023fractal}.

We obtain the stability phase diagram by computing the decorrelator, where $\mathbf m_i(0) = \{0,0,1/2\}$ and $\mathbf m_i^{\prime}(0) = \frac{1}{2}\{\sin(\theta_i) \cos(\phi_i),\sin(\theta_i) \sin(\phi_i),\cos(\theta_i)\}$, where $\theta_1=\theta_2 = \theta$ is small,
and $\phi_1, \phi_2$ are uniformly sampled from $[0, 2\pi)$. Finally, we take a time average of $D(nT)$ after discarding the initial transient and obtain $\overline{D}$. Our results shown in Fig.~\ref{fig:1}(c) demonstrate that there are wide parameter regimes where the time-averaged decorrelator $\overline{D}$ is small ($\sim 0$), leading to the emergence of stable dynamical regimes that host both coexisting and global DTCs. We note that this phase diagram exhibits a rich symmetric structure. A detailed discussion of these symmetries is presented in the Supplemental Material~\cite{suppmat}.

We now investigate the nature of the DTCs that emerge in this 
system by analyzing the behavior of ${\bf m}_1, {\bf m}_2$ for 
various cuts in the stability phase diagram.

{\emph{Coexisting and global DTC phases:}} A promising place to start our analysis is to consider a cut through the decorrelator phase diagram at $h_1=0.5$. In this case, $\overline{D}\sim 0$ for a wide parameter regime, thereby pointing to the presence of dynamical phases where inequivalent temporal orders emerge in the two sub-regions. To characterize these phases, we analyze the dynamics of $m_1^z, m_2^z$, as well as the Fourier transform of $m^z=m_1^z+m_2^z$, $\tilde{m}^z (\nu)$. The resulting phase diagram (Fig.~\ref{fig:2}(a1)) reveals the presence of a rich landscape of non-ergodic phases where a period-doubling DTC ($S_1$) coexists with a higher-order (HO)-DTC. These coexisting DTCs are characterized by two sharp and robust peaks in $\tilde{m}^z (\nu)$, and are dynamically manifested by distinct oscillation periods of the subregion magnetization (Fig.~\ref{fig:2}(b1)). The system also hosts other dynamical phases, where a period-doubling DTC coexists with a non-ergodic oscillatory phase (OP) (Fig.~\ref{fig:2}(b2)), characterized by a subharmonic response, whose dominant peak changes continuously with changing $h_2$~\cite{pizzi2019period}. Moreover, when $h_2$ is small, a chimera DTC~\cite{sakurai2021chimera,rahaman2024time} emerges in this system, where a period-doubling DTC coexists with a Floquet-synchronized state (Fig.~\ref{fig:2}(b3)). Finally, when $h_2 \approx 0.5$, the system realizes a global DTC with a robust period-doubling response in both sub-regions, consistent with the uniform drive limit~\cite{pizzi2021higher}. Thus, this cut provides examples of several kinds of coexisting dynamical phases. Notably, however, in all of these phases one of the sub-regions is always in a period-doubling DTC phase. It is therefore crucial to investigate other cuts where only higher-order DTCs coexist in different regions of the system.

To this end, we consider a cut at $h_1=0.16$, where tuning $h_2$ enables the system to transition between chaotic and non-ergodic regimes. In this case, the Fourier transform of $m^z$ (Fig.~\ref{fig:2}(a2)) reveals the existence of dynamical phases where two HO-DTCs coexist in different sub-regions. A robust example of this phenomenon is the coexistence of period-$3T$ and period-$6T$ DTCs (Fig.~\ref{fig:2}(b4)). Crucially, although this coexisting DTC and the global period-$6T$ DTC have the same overall response period, their Fourier response is qualitatively distinct; the former is characterized by two sharp peaks, while the latter exhibits a single sharp peak. Furthermore, it is not possible to deform the coexisting DTC to a global DTC with the same overall period (in this case, $6T$) by tuning $h_2$ without encountering a chaotic regime, thereby demonstrating that the coexisting $3T\!-\!6T$ DTC phase is distinct from the global $6T$-DTC phase. In addition to coexisting DTCs, the system hosts a family of global DTC phases ($S_2$) that emerge when $h_2$ is near $h_1$. Interestingly, in this regime, it is possible to steer the system between different global DTC phases by tuning the transverse field. Our analysis so far has established the existence of several distinct classes of coexisting DTCs in the thermodynamic limit. We emphasize that this coexistence emerges despite the global nature of the interactions. We now proceed to examine whether this coexistence phenomenology survives in the presence of quantum fluctuations in experimentally relevant finite-size systems.

{\emph{Quantum dynamics:}} To examine the fate of the DTC coexistence in the presence of quantum fluctuations, a natural first step is to compute the fidelity out-of-time-ordered correlator (FOTOC), which serves as the quantum analog of the decorrelator~\cite{LewisSwan2019}:
\begin{equation}
\begin{split}
\mathcal{F}_G(t)
&=1-\left|\langle \psi(t) |e^{i\epsilon G}| \psi(t) \rangle\right|^2 \\
&\approx
\epsilon^2 \left(\langle G^2(t)\rangle-\langle G(t)\rangle^2\right)
\equiv \epsilon^2 F_G ,
\end{split}
\end{equation}
where $G$ is usually a collective-spin operator; non-ergodic and chaotic regimes are characterized by a low ($\sim 0$) and high value of $F_G$ ($\sim 1$), respectively. Due to the presence of two sub-regions in our system, a natural version of $F_G$ is given by~\cite{mondal2021dynamical}:
\begin{equation}
F_i(t)=
\sum \limits_{\alpha=x,y,z}
\left[
\langle (\hat S_i^{\alpha})^2 \rangle
-
\langle \hat S_i^{\alpha} \rangle^2
\right]/ \langle \vec{S_i}. \vec{S_i}\rangle.
\label{eq:FOTOC}
\end{equation}
We characterize the dynamical phases by computing the region-averaged FOTOC, $F= (F_1+F_2)/2$.

\begin{figure}[t]
\centering
\includegraphics[width=0.94\linewidth]{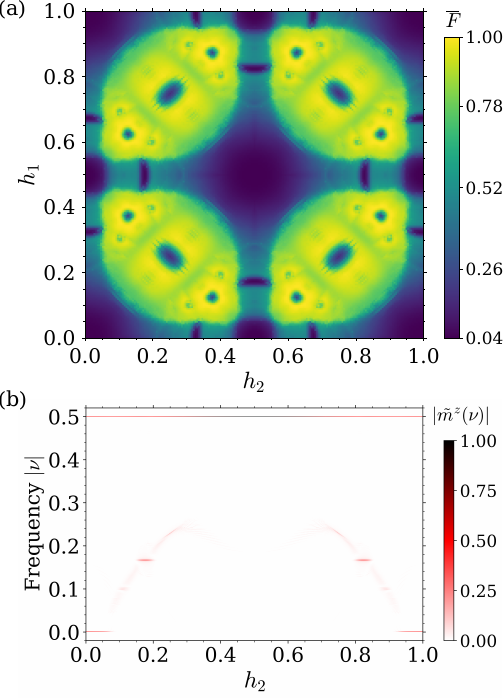}
\caption{{\bf Coexistence phenomenology for finite-size systems}
(a) The time-averaged stroboscopic FOTOC
$\overline{F} = (F_1 +F_2)/2$ ($F_i$ is defined in Eq.~\ref{eq:FOTOC}) for interaction strength $J=0.5$ for $N=100$, where the time-averaging has been performed over $1000$ drive periods from $n=1000$ to $n=2000$. The behavior of $\overline{F}$ closely mirrors the phase diagram of the time-averaged decorrelator, $\overline{D}$ (Fig.~\ref{fig:1}(c)). Notably, the dominant non-ergodic regimes persist in the presence of quantum fluctuations, even though some global and coexisting HO-DTCs are suppressed.
(b) Density plot of the Fourier spectrum for the $h_1=0.5$ cut, when $N=100$. The principal semiclassical dynamical regimes survive in the presence of quantum fluctuations, including coexisting DTCs, coexistence of DTCs with an oscillatory phase, and the chimera DTC, demonstrating that these coexisting phases are not an artifact of the semiclassical thermodynamic limit.}
\label{fig:3}
\end{figure}

Our results, shown in Fig.~\ref{fig:3}(a) for $N=100$, demonstrate that the behavior of the FOTOC closely mirrors the semiclassical decorrelator, and the minima of the time-averaged FOTOC, $\overline{F}$, coincide with those of $\overline{D}$. Notably, the dominant non-ergodic regimes observed in the thermodynamic limit survive in the presence of quantum fluctuations, even though some of the global and coexisting HO-DTC regions are suppressed. We note that this suppression is a finite-size effect; for $h_1=h_2$, the semiclassically predicted global DTCs emerge at larger system-sizes $(N \sim 1000)$~\cite{suppmat}. This observation suggests that additional coexisting phases may emerge with increasing $N$, though a systematic study of this phenomenon is beyond the scope of this work.

Despite the reduction of the non-ergodic regimes due to quantum fluctuations, the principal coexistence phenomenology persists for $N=100$. We demonstrate this by taking a representative cut through the parameter space at $h_1=0.5$. The Fourier response (Fig.~\ref{fig:3}(b)) shows that this system hosts the primary coexisting dynamical regimes observed semiclassically, including coexisting DTCs, chimera DTCs, as well as the coexistence of period-doubling DTCs with oscillatory phases. Furthermore, the coexistence of HO-DTC regimes can also be observed for this system-size, as discussed in the supplemental material~\cite{suppmat}. We conclude that the coexistence of distinct kinds of DTCs in the globally coupled LMG model is not an artifact of the thermodynamic limit. It remains observable in the presence of quantum fluctuations in relatively modest-sized quantum systems ($N=100$). This makes our proposal amenable to near-term experimental realizations with trapped ions, where the LMG model has been realized with $N \sim 100 $~\cite{bohnet2016quantum,garttner2017measuring}.

{\emph{Conclusion and Outlook:}} In this work, we have demonstrated that a minimal spatially non-uniform driving protocol characterized by two different transverse driving fields, $h_1$ and $h_2$, can be harnessed to realize the \emph{coexistence of distinct dynamical phases} in collective quantum systems. Notably, despite the global nature of interactions, the system self-organizes into sub-regions that break the discrete time-translation symmetry of the Hamiltonian in inequivalent ways, giving rise to robust and long-lived subharmonic responses with different periodicities.

A key result of our work is the emergence of non-ergodic phases where a DTC coexists with either another distinct DTC, an oscillatory phase, or a Floquet-synchronized phase (chimera DTC) in an infinite-range interacting collective spin system. Crucially, the coexisting DTCs are characterized by two robust subharmonic peaks, due to distinct kinds of symmetry breaking in two sub-regions. These coexisting DTCs are thus distinct from DTCs with a single global period. Furthermore, a tunable family of global DTC phases emerges when $h_1$ and $h_2$ are proximate. Due to the collective nature of the system, we establish the existence of these coexisting phases directly in the thermodynamic limit. Importantly, we demonstrate that the coexistence phenomenology is not an artifact of the semiclassical thermodynamic limit, and it persists in the presence of quantum fluctuations in modest-sized quantum systems, accessible in state-of-the-art trapped-ion experiments with $N \sim 100$ ions. We note that measuring the Fourier response of the global magnetization provides a concrete route for the experimental certification of these coexisting phases. Taken together, our work introduces a route to engineer non-equilibrium phase coexistence and realize novel forms of spatiotemporal order in driven collective systems. 

There are several natural avenues for future work. An important first step would be to develop an analytical framework to predict the classes of coexisting dynamical phases that can be realized by spatially non-uniform driving. In the thermodynamic limit, a natural route toward building such a framework may be provided by bifurcation theory. Furthermore, it would be interesting to generalize this protocol to quasiperiodic and other forms of structured aperiodic driving and examine the nature of coexisting phases that emerge beyond the Floquet paradigm. It would also be interesting to investigate the effect of dissipation on many-body systems subjected to a spatially non-uniform drive. Finally, we note that the boundary between the sub-regions that host distinct kinds of temporal orders may be viewed as a temporal analog of a spatial domain wall. It would be intriguing to explore the possibility of creating other kinds of temporal topological defects through spatially structured driving.

\begin{acknowledgments}
{\emph{Acknowledgements:}} The authors acknowledge institutional support from Harish-Chandra Research Institute (HRI) and Homi Bhabha National Institute (HBNI). SC acknowledges support from the Department of Science and Technology (DST), India, through SERB Project No.~SRG/2023/002730 and through Project No.~DST/FFT/NQM/QSM/2024/3. SM acknowledges the high-performance computing facility at HRI for computational resources.
\end{acknowledgments}

\bibliography{references}

\onecolumngrid
\section*{End Matter}

\setcounter{figure}{0}
\renewcommand{\thefigure}{A\arabic{figure}}

\setcounter{equation}{0}
\renewcommand{\theequation}{A.\arabic{equation}}

\setcounter{table}{0}
\renewcommand{\thetable}{A.\arabic{table}}

\setcounter{section}{0}
\renewcommand{\thesection}{A.\Roman{section}}

\renewcommand{\thesubsection}{A.\Roman{section}.\Alph{subsection}}

\makeatletter
\renewcommand*{\p@subsection}{}
\makeatother

\renewcommand{\thesubsubsection}{E.\Roman{section}.\Alph{subsection}-\arabic{subsubsection}}

\makeatletter
\renewcommand*{\p@subsubsection}{}  
\makeatother

\twocolumngrid

In this End Matter, we derive the semiclassical equations of motion that can describe the dynamics of the system in the thermodynamic limit. To this end, we introduce the
collective-spin operators for the two sub-regions,

\begin{equation*}
\hat S_1^{\alpha}
=
\frac{1}{2}\sum_{j=1}^{N/2}\sigma_j^\alpha,
\qquad
\hat S_2^{\alpha}
=
\frac{1}{2}\sum_{j=N/2+1}^{N}\sigma_j^\alpha,
\qquad
\alpha=x,y,z,
\end{equation*}
where
$\boldsymbol{\sigma}_j=(\sigma_j^x,\sigma_j^y,\sigma_j^z)$ are Pauli matrices.
The corresponding subregion magnetizations are
\begin{equation*}
m_i^{\alpha}
=
\frac{2}{N}\langle \hat S_i^{\alpha}\rangle,
\qquad
i=1,2.
\end{equation*}

In terms of the collective-spin operators, Eq.~(\ref{eq:drive}) becomes
\begin{align}
H_1
&=
\frac{8J}{NT}
\left(\hat S_1^z+\hat S_2^z\right)^2,
\\
H_2
&=
-\frac{4\pi}{T}
\left(h_1\hat S_1^x+h_2\hat S_2^x\right).
\end{align}

We now note that due to the infinite-range nature of the interactions, mean-field theory becomes exact in the thermodynamic limit. Consequently, during the first half of the time evolution ($0<t\le T/2$), the sub-region magnetizations evolve according to:
\begin{align*}
\dot m_i^{x}
&=
-\frac{8J}{T}\,
m_i^{y}
\left(m_1^{z}+m_2^{z}\right),
\\
\dot m_i^{y}
&=
\frac{8J}{T}\,
m_i^{x}
\left(m_1^{z}+m_2^{z}\right),
\\
\dot m_i^{z}
&= 0.
\end{align*}

During the next half of the time-evolution ($T/2<t\le T$), the two regions evolve independently
under the transverse fields:
\begin{align*}
\dot m_i^{x}
&=
0,
\\
\dot m_i^{y}
&=
\frac{4\pi h_i}{T}\,
m_i^{z},
\\
\dot m_i^{z}
&=
-\frac{4\pi h_i}{T}\,
m_i^{y}.
\end{align*}

These equations can then be integrated exactly over one Floquet period leading to:

\begin{widetext}
\begin{align}
m_i^x \bigg((n+1) T\bigg)
&=
m_i^x (nT) \cos\Phi
-
m_i^y (nT) \sin\Phi,
\nonumber\\
m_i^y \bigg((n+1) T\bigg)
&=
\left(
m_i^y (nT) \cos\Phi
+
m_i^x (nT)\sin\Phi
\right)\cos\theta_i
+
m_i^z (nT) \sin\theta_i,
\nonumber\\
m_i^z \bigg((n+1) T\bigg)
&=
-\left(
m_i^y (nT)\cos\Phi
+
m_i^x (nT) \sin\Phi
\right)\sin\theta_i
+
m_i^z (nT) \cos\theta_i,
\label{eq:floquet_map}
\end{align}
\end{widetext}
where $\Phi=4J(m_1^z+m_2^z)$ and $\theta_i=2\pi h_i$.

We then employ Eq.~(\ref{eq:floquet_map}) to obtain the regional magnetizations
$m_1^z$ and $m_2^z$, as well as the total magnetization
$m_z=m_1^z+m_2^z$, at stroboscopic times.

\newpage
\clearpage
\begin{widetext}
\renewcommand{\thepage}{S\arabic{page}} 
\renewcommand{\thesection}{S\arabic{section}}  
\renewcommand{\thetable}{S\arabic{table}}  
\renewcommand{\thefigure}{S\arabic{figure}}
\renewcommand{\theequation}{S\arabic{equation}}
\setcounter{figure}{0}
\setcounter{page}{1}
\setcounter{equation}{0}
\begin{center}
\textbf{\Large{Supplemental Material for: `Coexistence of inequivalent time-crystalline orders in a Floquet collective spin system'}}
\end{center}
This Supplemental Material provides technical details supporting the 
results presented in the main text. The first section describes 
the semiclassical decorrelator and its use as a diagnostic for 
distinguishing DTCs and other non-ergodic phases from chaotic ones. 
The next section presents regional Fourier spectra and 
additional results on the time-evolution of the system in the thermodynamic 
limit. The parameter-space symmetries 
of the decorrelator are presented in the following section. Finally, 
details about the finite-size quantum calculations, 
including the exact diagonalization setup, quantum DFT spectra, and the 
system-size dependence of the FOTOC in the last section. For the semiclassical analysis 
presented, we employ 
the equations of motion derived in the End Matter of the main text.

\section{Semiclassical Decorrelator Analysis}
\label{sec:stability}

\begin{figure}[b]
\centering
\includegraphics[width=0.95\linewidth]{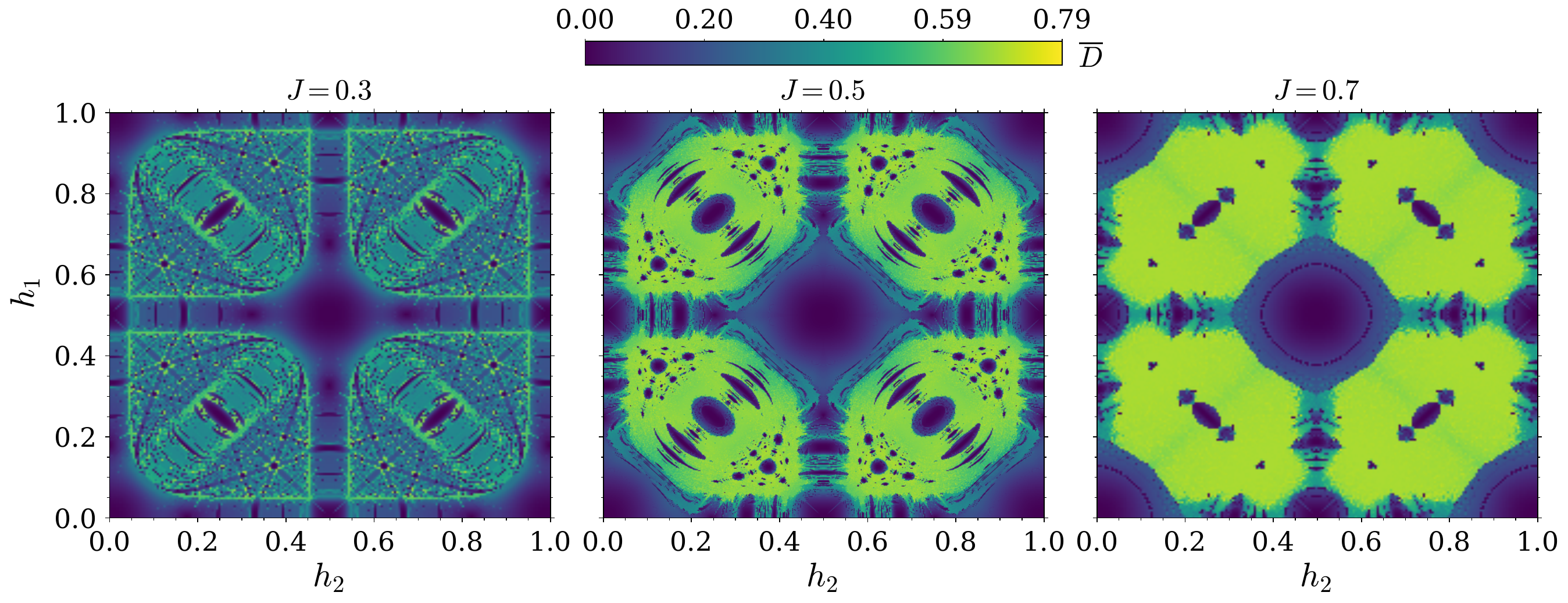}
\caption{
Long-time-averaged decorrelator $\overline{D}$ 
for interaction strengths (a) $J=0.3$, (b) $J=0.5$, and (c) $J=0.7$. 
The decorrelator is averaged over $1600$ uniformly sampled values of the azimuthal angles $\phi_1,\phi_2\in[0,2\pi)$, 
and subsequently averaged over $1000 - 2000$ Floquet periods. The chaotic regimes grow progressively with larger $J$.
}
\label{fig:DecNU}
\end{figure}

In this section, we provide further details about the behavior of the decorrelator in the thermodynamic limit. As noted in the main text, the decorrelator is the semiclassical analog of the out-of-time-ordered correlator (OTOC), and it measures the sensitivity to initial conditions. Consequently, the decorrelator, $D$ can be employed to distinguish non-ergodic phases ($D \sim 0$) from chaotic ones ($D \sim 1/\sqrt{2}$).  The decorrelator, $D$ after $n$ drive periods is defined as
\begin{equation}
D(nT)=
\sqrt{\frac{1}{2}\sum_{i=1}^{2}
\left|\mathbf m_i(nT)-\mathbf m^{\prime}_i(nT)\right|^2},
\label{eq:decorrelator}
\end{equation}
where 
$\mathbf m_i(nT)$ and $\mathbf m^{\prime}_i(nT)$ denote the magnetization after $n$ drive periods, for two initially close copies of the initial state~\cite{pizzi2021higher}. As noted in the main text, we obtain the stability phase diagram by considering $\mathbf m_i(0) = \{0,0,1/2\}$ and $\mathbf m_i^{\prime}(0) = \frac{1}{2}\{\sin(\theta_i) \cos(\phi_i),\sin(\theta_i) \sin(\phi_i),\cos(\theta_i)\}$, where $\theta_1=\theta_2 = \theta$ is small, and $\phi_1, \phi_2$ are uniformly sampled from $[0, 2\pi)$. Finally, we compute the time-averaged $D(nT)$, $\overline{D}$ at long times:
\begin{equation}
\overline{D}=
\frac{1}{n_f+1}\sum_{n=n_i}^{n_i+n_f}\overline{D}(nT),
\end{equation}
where $n_i$ is chosen to be sufficiently large to ensure that the initial transient has been discarded. 

Figure~\ref{fig:DecNU} shows the stability phase diagram in the $(h_1,h_2)$ plane for $J=0.3$, $0.5$, and $0.7$. At weak coupling ($J=0.3$), the stable and chaotic 
regions are only weakly separated. Intermediate coupling ($J=0.5$) yields a  rich landscape of well-defined stability islands, corresponding to the diverse  dynamical phases discussed in the main text, and is therefore adopted as the representative value throughout. Stronger $J$ increases the non-linearity ($J=0.7$ in Fig.~\ref{fig:DecNU}), thereby leading to a suppression of non-ergodic stable regions, and enhancement of the chaotic regions of the phase diagram.

\section{Semiclassical Spectral and Dynamical Diagnostics}
\label{sec:semiclassical}

\begin{figure}[t]
\centering
\includegraphics[width=1.0\linewidth]{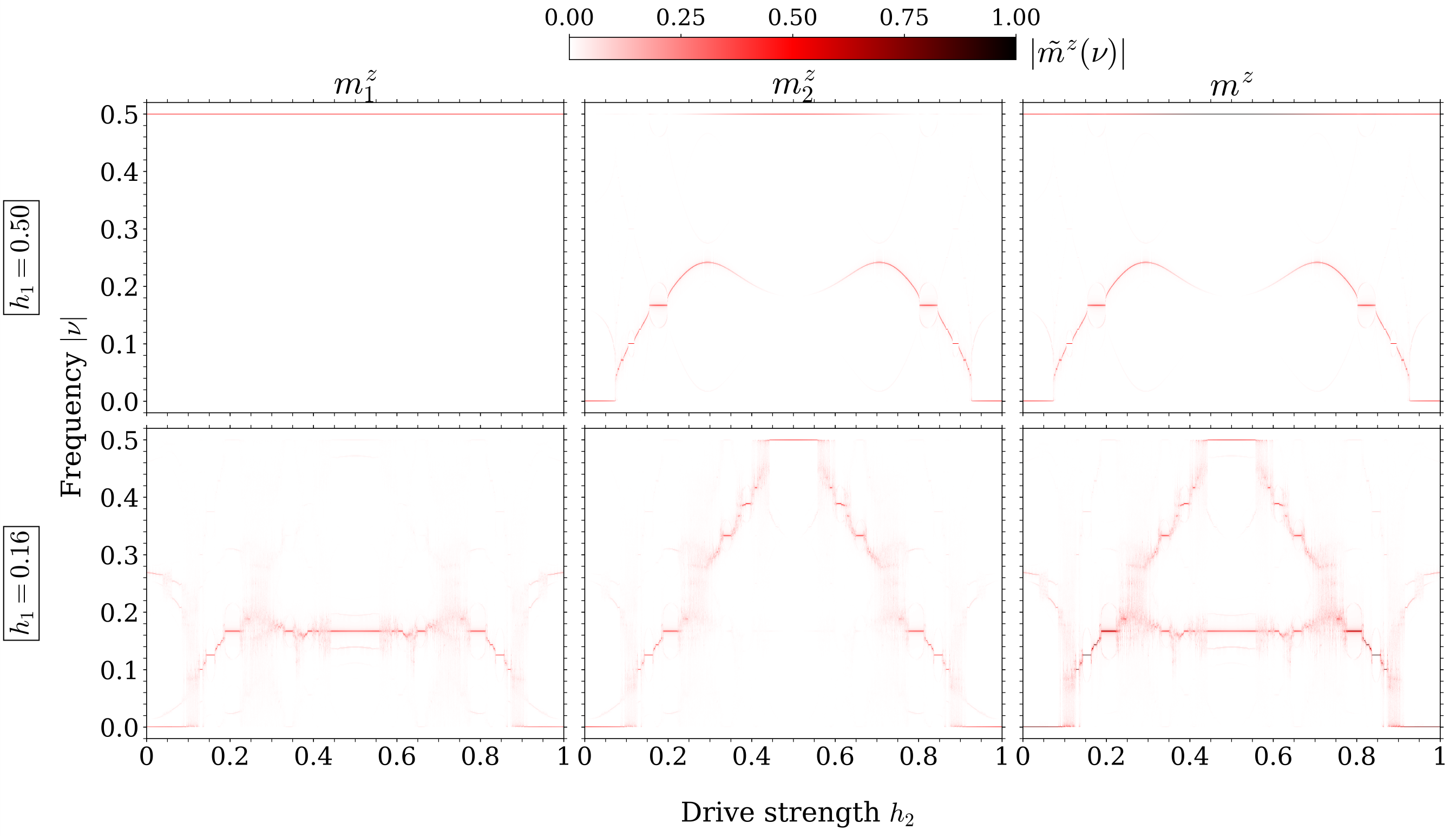}
\caption{
DFT density plots of the regional magnetizations $m_1^{z}$ and
$m_2^{z}$, together with the total magnetization $m_z$, for
$h_1=0.50$ and $h_1=0.16$. Inequivalent coexisting DTC orders in regions 1 and 2 are manifested by distinct sharp peaks in $m_1^z$ and $m_2^z$ and two robust peaks in $m^z$.
}
\label{fig:fft_sc}
\end{figure}

In this section, we present discrete Fourier transform (DFT) density plots of the stroboscopic magnetization  resolved by spatial subregion. While the total magnetization captures the dominant global response, resolving $m_1^{z}$ and $m_2^{z}$ individually is essential to identify which subharmonic response belongs to which region, providing direct 
evidence for coexistence of inequivalent time-crystalline orders in different subregions. For a given set of driving parameters $(h_1,h_2)$, the semiclassical equations of motion are evolved for $1000$ drive periods. The stroboscopic time series $m_i^{z}(nT)$ ($i=1,2$), together with the total magnetization $m^z(nT)=m_1^{z}(nT)+m_2^{z}(nT)$, are Fourier transformed to obtain the DFT density plots as a function of $h_2$ at fixed $h_1$.

Figure~\ref{fig:fft_sc} shows the regional and total DFT density plots for
the two representative cuts discussed in the main text. For $h_1=0.50$,
region~1 always hosts a period-$2T$-DTC. On the other hand, region~2 remains Floquet-synchronized
(period-$T$) for $h_2\lesssim 0.07$, realizing a chimera-like dynamical phase in this regime. Increasing $h_2$ further leads to the coexistence of the period-$2T$-DTC in
region~1 with a tunable set of higher-order (HO)- DTCs in region 2.  As shown in the top panel of Fig.~\ref{fig:fft_sc}, increasing $h_2$ leads to the successive appearance of HO-DTCs with periods of $18T$, $14T$, $12T$, $10T$, $8T$, and
$6T$ respectively in region~2. Increasing $h_2$ further leads to the formation of a non-ergodic oscillatory phase in region~2. Finally, the system evolves into a global period-doubling DTC phase. For $h_1=0.16$ and $h_2\simeq0.5$, region~1 exhibits a robust $6$-DTC,
while region~2 simultaneously supports a $2$-DTC. Around $h_2=0.34$,
coexistence between period-$6T$ and period-$3T$ DTCs is observed. As $h_2$
is reduced further, the system synchronizes into a family of global DTC
phases with periods $6T$, $7T$, $8T$, $9T$, and $10T$, each characterized
by a single dominant DFT peak in contrast to the two peaks seen in the
coexisting DTC phases. Crucially, these results establish that the presence of two robust peaks in the DFT spectrum implies that the system hosts inequivalent DTC orders in different regions. 

We complement this discussion on the Fourier response with an analysis of the magnetization dynamics in Fig.~\ref{fig:time_domain} for some representative parameter points. For
$h_1=0.5$, the coexistence of a period-$2T$ DTC in region~1 with a period-$14T$
($h_2=0.09$) and period-$10T$ ($h_2=0.11$) DTCs in region~2 is shown. For
$h_1=0.16$, global DTCs with period-$8T$ ($h_2=0.153$) and period-$7T$
($h_2=0.173$) are displayed. These results, in conjunction with Fig.~2 (b1)-(b4) of the main text provide a wide set of archetypal examples of the dynamical phases hosted by this system.

\begin{figure}[t]
\centering
\includegraphics[width=\linewidth]{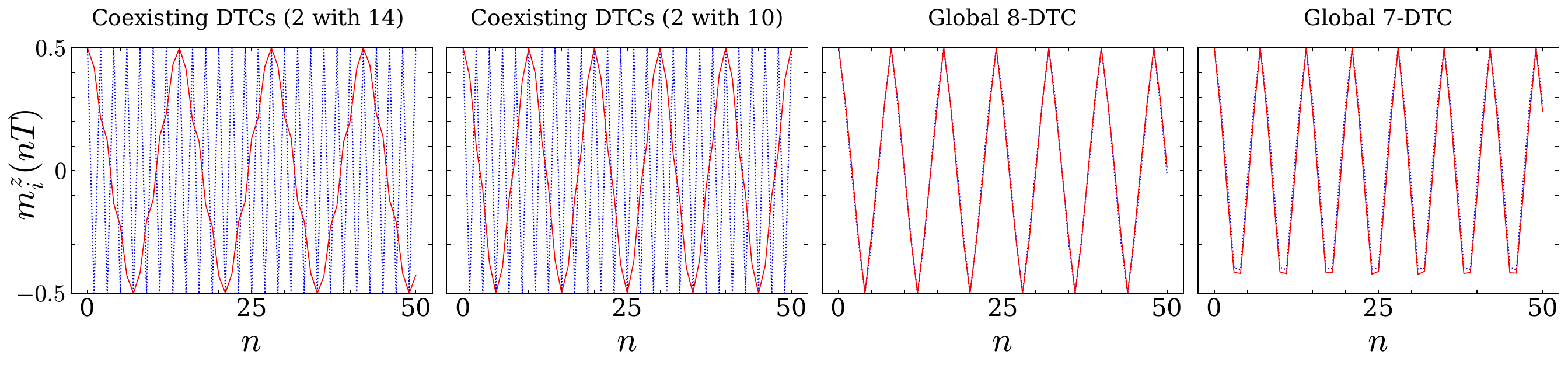}
\caption{
Representative semiclassical regional magnetization dynamics for
$m_1^{z}(nT)$ (blue) and $m_2^{z}(nT)$ (red)
for parameter points selected from distinct regions of the decorrelator phase diagram. For $h_1=0.5$, the figure shows coexistence between a period-$2T$ DTC in region~1 and HO subharmonic oscillations in region~2, including period-$14T$ at $h_2=0.09$ and period-$10T$ at $h_2=0.11$. For $h_1=0.16$, HO global DTCs with period-$8T$
at $h_2=0.153$ and period-$7T$ at $h_2=0.173$ are shown. These results supplement the regional magnetization dynamics presented in Fig.~2 (b1)-(b4) of the main text.
}
\label{fig:time_domain}
\end{figure}

\section{Symmetries of the decorrelator}
\label{sec:symmetry}

The decorrelator phase diagram presented in Fig.~1(c) of the main text exhibits a rich symmetric structure. In this section, we elucidate the origin of these symmetries.

\subsection{Sub-region exchange symmetry}
We first demonstrate that the decorrelator exhibits a sub-region exchange symmetry. To this end, we note that the Floquet operator corresponding to Eq.~(A.1) and~(A.2) in the End Matter is
\begin{equation}
U(h_1,h_2)
=
e^{i2\pi(h_1\hat S_1^x+h_2\hat S_2^x)}
e^{-i\frac{4J}{N}(\hat S_1^z+\hat S_2^z)^2}.
\label{eq:floquet_symmetry}
\end{equation}
The Floquet unitary satisfies
\begin{equation}
U(h_1,h_2)
=
X\,U(h_2,h_1)\,X^\dagger,
\end{equation}
where $X$ is the unitary operator that permutes the two subregions.
This follows immediately from the invariance of
$(\hat S_1^z+\hat S_2^z)^2$ under subregion exchange and the relation
\begin{equation}
X(h_1\hat S_1^x+h_2\hat S_2^x)X^\dagger
=
h_2\hat S_1^x+h_1\hat S_2^x.
\end{equation}
In the semiclassical description, the exchange operator $X$
simply interchanges the collective-spin vectors $\mathbf m_1$ and $\mathbf m_2$. The decorrelator is therefore symmetric under
$(h_1,h_2)\leftrightarrow(h_2,h_1)$, thereby explaining the reflection about
the diagonal $h_1=h_2$. Since the decorrelator depends only on the distance between nearby trajectories, it is unchanged under this relabeling.

\subsection{Floquet $\mathbb{Z}_2$ Symmetry}

We next explore the signature of the $\mathbb{Z}_2$ symmetry of the Floquet operator on the decorrelator. To this end, we note that the $\pi$-rotation operators about the $x$-axis,
\begin{equation}
P_1=e^{-i\pi\hat S_1^x},
\qquad
P_2=e^{-i\pi\hat S_2^x},
\qquad
P=P_1P_2.
\end{equation}
transform $\hat S_i^z\rightarrow-\hat S_i^z$ while leaving
$\hat S_i^x$ invariant. Consequently, we obtain
\begin{equation}
P(\hat S_1^z+\hat S_2^z)^2P^\dagger = (\hat S_1^z+\hat S_2^z)^2,
\end{equation}
and
\begin{equation}
[P,U(h_1,h_2)]=0.
\end{equation}
This is the $\mathbb{Z}_2$ symmetry of the Floquet operator.

Using
\begin{equation}
e^{i2\pi(\frac12)\hat S_i^x}
=
P_i^\dagger,
\end{equation}
the Floquet operator at shifted parameters becomes
\begin{align}
U\!\left(h_1+\tfrac{1}{2},h_2+\tfrac{1}{2}\right)
&=
e^{i2\pi[(h_1+\frac{1}{2})\hat S_1^x
+(h_2+\frac{1}{2})\hat S_2^x]}
e^{-i\frac{4J}{N}(\hat S_1^z+\hat S_2^z)^2}
\nonumber\\
&=
P^\dagger U(h_1,h_2).
\end{align}

Since $[P,U(h_1,h_2)]=0$, it follows that
\begin{equation}
U\!\left(h_1+\tfrac{1}{2},h_2+\tfrac{1}{2}\right)^n
=
(P^\dagger)^n U(h_1,h_2)^n.
\end{equation}
From the action of $P$ on the spin operators, it follows that the
stroboscopic magnetizations transform according to
\begin{equation}
\bigg(m_i^x (n),m_i^y (n),m_i^z (n)\bigg)
\rightarrow
\bigg(m_i^x (n),(- 1)^n m_i^y (n),(- 1)^n m_i^z(n) \bigg).
\end{equation}
Since this transformation applies to both the reference and perturbed trajectories, the decorrelator stays unchanged:
\begin{equation}
D(h_1,h_2)
= D \!\left(h_1+\tfrac{1}{2},h_2+\tfrac{1}{2}\right).
\label{eq:Dshift}
\end{equation}

\subsection{Reflection Symmetry}

We now note that for initial states that are eigenstates of the $\hat{S}^z = \hat{S}_1^z + \hat{S}_2^z$ operator, an additional reflection symmetry is present in the decorrelator phase diagram: $\overline{D}(h_1,h_2) = \overline{D} \!\left(\tfrac{1}{2}-h_1,\tfrac{1}{2}-h_2\right)$. To elucidate the origin of this symmetry, we define the joint $\pi$-rotation about the $z$-axis,
\begin{equation}
Q=e^{-i\pi(\hat S_1^z+\hat S_2^z)}.
\end{equation}
Under this transformation,
\begin{equation}
Q\hat S_i^xQ^\dagger=-\hat S_i^x,
\qquad
Q\hat S_i^zQ^\dagger=\hat S_i^z,
\end{equation}
which gives
\begin{equation}
Q\,U(h_1,h_2)\,Q^\dagger
=
U(-h_1,-h_2).
\label{eq:qsym}
\end{equation}
Consequently, the stroboscopic magnetizations for any $\hat{S}^z$ eigenstate transform as:
\begin{equation}
\bigg(m_i^x (n),m_i^y (n),m_i^z (n)\bigg)
\rightarrow
\bigg(-m_i^x, -m_i^y (n), m_i^z(n) \bigg),
\end{equation}

Consequently, the decorrelator satisfies
\begin{equation}
D(h_1,h_2) = D(-h_1,-h_2).
\label{eq:Dqsym}
\end{equation}
Combining Eq.~(\ref{eq:Dqsym}) with the $\mathbb{Z}_2$ shift
Eq.~(\ref{eq:Dshift}) then yields 
\begin{equation}
\overline{D}(h_1,h_2)
=
\overline{D}
\!\left(\tfrac{1}{2}-h_1,\tfrac{1}{2}-h_2\right).
\end{equation}
An analogous symmetry also emerges in the uniformly
driven Floquet spin model~\cite{giachetti2023fractal}.

The additional symmetries of the decorrelator phase diagram, such as $(h_1,h_2)\rightarrow(1-h_1,h_2)$, $(h_1,h_2)\rightarrow(h_1,1-h_2)$,
and $(h_1,h_2)\rightarrow(1-h_1,1-h_2)$, follow directly from the unit
periodicity of the drive amplitudes. Finally, we note that these derivations establish the symmetries at the level of the quantum Floquet operator, and are valid for all system sizes, including the thermodynamic limit, where the semiclassical mean-field description becomes exact.

\section{Finite-Size Quantum Dynamics}
\label{sec:quantum}

\begin{figure}[t]
\centering
\includegraphics[width=0.85\linewidth]{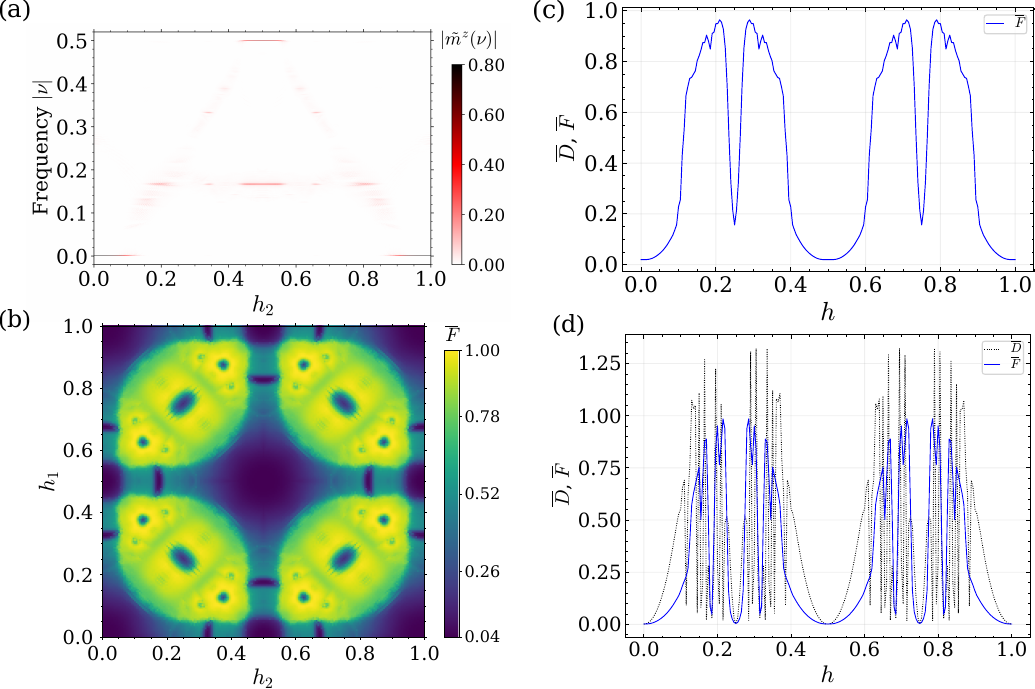}
\caption{
(a) Quantum DFT spectrum of the total magnetization $m^z(nT)$
for a representative parameter cut at $h_1=0.16$ and $N=100$.
Dominant subharmonic peaks associated with HO DTC regimes remain
visible despite finite-size quantum fluctuations.
(b) Long-time-averaged FOTOC $\overline{F}$ for $N=100$ and interaction strength $J=0.5$, plotted in the
$(h_1,h_2)$ plane and averaged over $1000-2000$ drive periods.
The dominant nonergodic regions track the minima of the
semiclassical decorrelator.
(c) Long-time-averaged FOTOC $\overline{F}$ for the uniformly driven
system ($h_1=h_2$) with $N=100$.
(d) Comparison between the long-time-averaged FOTOC $\overline{F}$
for $N=1000$ (blue) and the corresponding semiclassical decorrelator
$\overline{D}$ (black) in the uniform-drive case.
Several HO DTC regions become resolved at large system size,
demonstrating that the suppression of HO phases at $N=100$
is a finite-size effect.}
\label{fig:quantum}
\end{figure}

In this section, we provide details of the finite-size quantum calculations,  underpinning the results presented in the main text, including the  Hilbert-space reduction, quantum DFT spectra, and the FOTOC analysis.

\subsection{Exact diagonalization and Hilbert-space reduction}

We study the fully quantum dynamics of a finite system of $N=100$
spin-$1/2$ particles using exact diagonalization. Owing to the permutation
symmetry within each spatial subregion, the dynamics is restricted to the
symmetric collective-spin subspace of each half, reducing the Hilbert-space dimension from $2^N$ to $(N/2+1)^2$ and enabling numerically exact simulations.

The collective-spin operators $\hat S_i^\alpha=\frac{1}{2}\sum_{j\in i}\sigma_j^\alpha$ ($i=1,2$) are represented in the symmetric Dicke basis of each subregion using standard angular-momentum algebra. The Floquet operator is constructed as
\begin{equation}
U_F=e^{-iH_2T/2}e^{-iH_1T/2},
\end{equation}
and iterated over many drive periods within the reduced Hilbert space.
The normalized regional magnetizations $m_1^z$ and $m_2^z$, together with
the total magnetization $m^z=m_1^z+m_2^z$, are evaluated stroboscopically
and analyzed using the same DFT procedure employed in the semiclassical
calculations.

\subsection{Quantum DFT spectra}

In the main text, we presented results for the coexisting DTC phases that emerge for the $h_1=0.5$ cut. In this section, we provide evidence for the coexistence of inequivalent HO-DTCs in the presence of quantum fluctuations. To this end, we take a cut at $h_1 = 0.16$ for $N=100$, and analyze the DFT spectrum. Our results are shown in Fig.~\ref{fig:quantum}(a). Notably, the dominant subharmonic peaks remain clearly visible, even at $N=100$, including signatures of the coexisting period-$3T$ and period-$6T$ DTC phases, though finite-size quantum fluctuations substantially suppress the spectral weight of HO responses. We have verified that the regional spectra of $m_1^{z}$ and $m_2^{z}$ separately exhibit the same coexistence structure; for conciseness, only the total-magnetization spectrum is presented here.

\subsection{Fidelity out-of-time-order correlator}

To characterize the finite-size quantum dynamics, we compute the
region-averaged FOTOC defined in the main text,
\begin{equation}
F_i(t)=
\sum \limits_{\alpha=x,y,z}
\left[
\langle (\hat S_i^{\alpha})^2 \rangle
-
\langle \hat S_i^{\alpha} \rangle^2
\right]/ \langle \vec{S_i}. \vec{S_i}\rangle,
\label{eq:FOTOC}
\end{equation}
which serves as the quantum analog of the semiclassical decorrelator~\cite{mondal2021dynamical}.
The quantity $\overline{F}$ denotes the corresponding long-time average
computed over stroboscopic evolution from $1000$ to $2000$ Floquet periods.

For the non-uniformly driven system ($h_1 \ne h_2$), exact quantum calculations are restricted to $N=100$, since the Hilbert-space dimension scales as $(N/2+1)^2$. The resulting $\overline{F}$ is shown in
Fig.~\ref{fig:quantum}(b), where the dominant minima track those
of the semiclassical decorrelator. However, several HO DTC regimes are substantially weakened relative to their semiclassical counterparts. 

To determine whether the suppression of HO DTC regimes reflects
genuine destruction of these phases or merely finite-size effects, we examine the spatially uniform-drive regime ($h_1 = h_2$), where the Hilbert space dimension scales as $(N+1)$ and system sizes up to $N=1000$ become accessible. For $N=100$, only the dominant period-$2T$ and period-$4T$ DTC regions are clearly resolved (Fig.~\ref{fig:quantum}(c)). Furthermore, small dips of $\overline{F}$ corresponding to period-$3T$ and $6T$ DTCs can also be observed. In contrast, the $N=1000$ results exhibit
multiple additional minima coinciding with the HO DTC phases
predicted semiclassically, and the large-$N$ FOTOC closely follows the
semiclassical decorrelator (Fig.~\ref{fig:quantum}(d)). These results establish that the suppression of HO DTC regions at $N=100$ is primarily a finite-size effect.
\end{widetext}

\end{document}